\documentclass{aa}  

\usepackage{txfonts}
\usepackage{color}
\usepackage{graphicx}
\usepackage{natbib}
\usepackage{amsmath}
\usepackage{url} 
\usepackage[breaklinks=true,pdfmenubar=true,pdftoolbar=true,colorlinks=true,citecolor=blue,linkcolor=blue,urlcolor=blue]{hyperref}
\bibpunct{(}{)}{;}{a}{}{,} 
\begin{document}

   \title{Spatio-kinematical model of the collimated molecular outflow in the water-fountain nebula IRAS~16342$-$3814}


   \author{D. Tafoya\inst{1,2}, G. Orosz\inst{3,4}, W.~H.~T. Vlemmings\inst{2}, R. Sahai\inst{5}, \and A. F. Pérez-Sánchez\inst{6}
          }

   \institute{Chile Observatory, National Astronomical Observatory of Japan, National Institutes of Natural Science, 
   	2-21-1 Osawa, Mitaka, Tokyo, 181-8588, Japan
   \and
   Department of Space, Earth and Environment, Chalmers University of Technology, Onsala Space Observatory, 
   439~92 Onsala, Sweden\\  \email{daniel.tafoya@chalmers.se}
    \and
    School of Natural Sciences, University of Tasmania, Private Bag 37, Hobart, Tasmania 7001, Australia
    \and
   Xinjiang Astronomical Observatory, Chinese Academy of Sciences, 150 Science 1-Street, Urumqi, Xinjiang 830011, China
    \and
   Jet Propulsion Laboratory, MS 183-900, California Institute of Technology, Pasadena, CA 91109, USA
    \and
   European Souther Observatory, Alonso de Córdova 3107, Vitacura, Casilla 19001, Santiago de Chile
             }


 
  \abstract
   {Water fountain nebulae are AGB and post-AGB objects that exhibit high-velocity outflows traced by water maser emission. 
   	Their study is important to understand the interaction between collimated jets and the circumstellar material that 
   	leads to the formation of bipolar/multi-polar morphologies in evolved stars.}
   {To describe the three-dimensional morphology and kinematics of the molecular gas of the water-fountain nebula 
   	IRAS~16342$-$3814.}
   {Retrieving data from the ALMA archive to analyse it using a simple spatio-kinematical model. Using the software SHAPE to 
   	construct a three-dimensional spatio-kinematical model of the molecular gas in IRAS~16342$-$3814. Reproducing the intensity 
   	distribution and position-velocity diagram of the CO emission from the ALMA observations to derive the 
   	morphology and velocity field of the gas. Using CO($J$=1$\rightarrow$0) data to support the physical interpretation of the model.}
   {A spatio-kinematical model that includes a high-velocity collimated outflow embedded within material expanding 
   	at relatively lower velocity reproduces the images and position-velocity diagrams from 
   	the observations. The derived morphology is in good agreement with previous results from IR and 
   	H$_{2}$O maser emission observations. The high-velocity collimated outflow exhibits deceleration across its length, while the velocity of 
   	the surrounding component increases with distance. The morphology of the emitting region; the velocity field and the 
   	mass of the gas as function of velocity are in excellent agreement with the properties predicted for a molecular outflow 
   	driven by a jet. The timescale of the molecular outflow is estimated to be $\sim$70-100~years. The scalar momentum 
   	carried by the outflow is much larger than it can be provided by the radiation of the central star. An 
    oscillating pattern was found associated to the high-velocity collimated outflow. The oscillation period of the pattern 
    is $T$$\approx$60-90~years and its opening angle is $\theta_{\rm op}$$\approx$2$^{\circ}$.}
   {The CO ($J$=3$\rightarrow$2) emission in IRAS~16342$-$3814 is interpreted in terms of a jet-driven molecular outflow expanding along 
   	an elongated region. The position-velocity diagram and the mass spectrum reveal a feature due to entrained material 
   	that is associated to the driving jet. This feature is not seen in other more evolved objects that exhibit more developed 
   	bipolar morphologies. It is likely that the jet in those objects has already disappeared since it is expected to last only for a 
   	couple of hundred years. This strengthens the idea that water fountain nebulae are undergoing a very short transition 
   	during which they develop the collimated outflows that shape the CSE. The oscillating pattern seen in the CO high-velocity	
   	outflow is interpreted as due to precession with a relatively small opening angle. The precession period is compatible with 
   	the period of the corkscrew pattern seen at IR wavelengths. We propose that the high-velocity molecular outflow traces 
   	the underlying primary jet that produces such pattern.}


\authorrunning{D. Tafoya et al.}
\titlerunning{Spatio-kinematical model of IRAS~16342$-$3814}
\maketitle
%

\section{Introduction}

There is a growing consensus that collimated jet-like outflows create cavities with dense walls in the slowly expanding 
circumstellar envelopes (CSE) of post-Asymptotic Giant Branch (post-AGB) stars. Once the central star becomes hot 
enough to ionise the circumstellar material, such structures are seen as beautiful and colourful bipolar or even multi-polar 
planetary nebulae (PNe) \citep{Sahai1998}. Water-fountain nebulae (wf-nebulae) are a reduced group of AGB and post-AGB 
objects that are thought to be experiencing the first manifestation of such collimated outflows \citep{Imai2007}. 
Therefore these objects are one of the best tools that we have available to probe and study the mechanisms of jet 
launching and collimation in evolved stars, as well as the interaction of the jet with the CSE.

IRAS~16342$-$3814 was the first object to be classified as a wf-nebula given the wide spread of its H$_{2}$O maser 
emission of $\sim$260~km~s$^{-1}$ \citep{Likkel1988}. IRAS~16342$-$3814 is one of the best studied wf-nebulae, 
observed over a wide range of wavelengths across the electromagnetic spectrum, from optical to radio  
\citep{Sahai1999,Sahai2005,Gledhill2012,Imai2012,Sahai2017}. The distribution and kinematics of the circumstellar 
material has also been studied via its H$_{2}$O and OH maser emission \citep{Claussen2009,Sahai1999}. 

\begin{figure*}
	\centering
	\includegraphics[angle=0,scale=0.6]{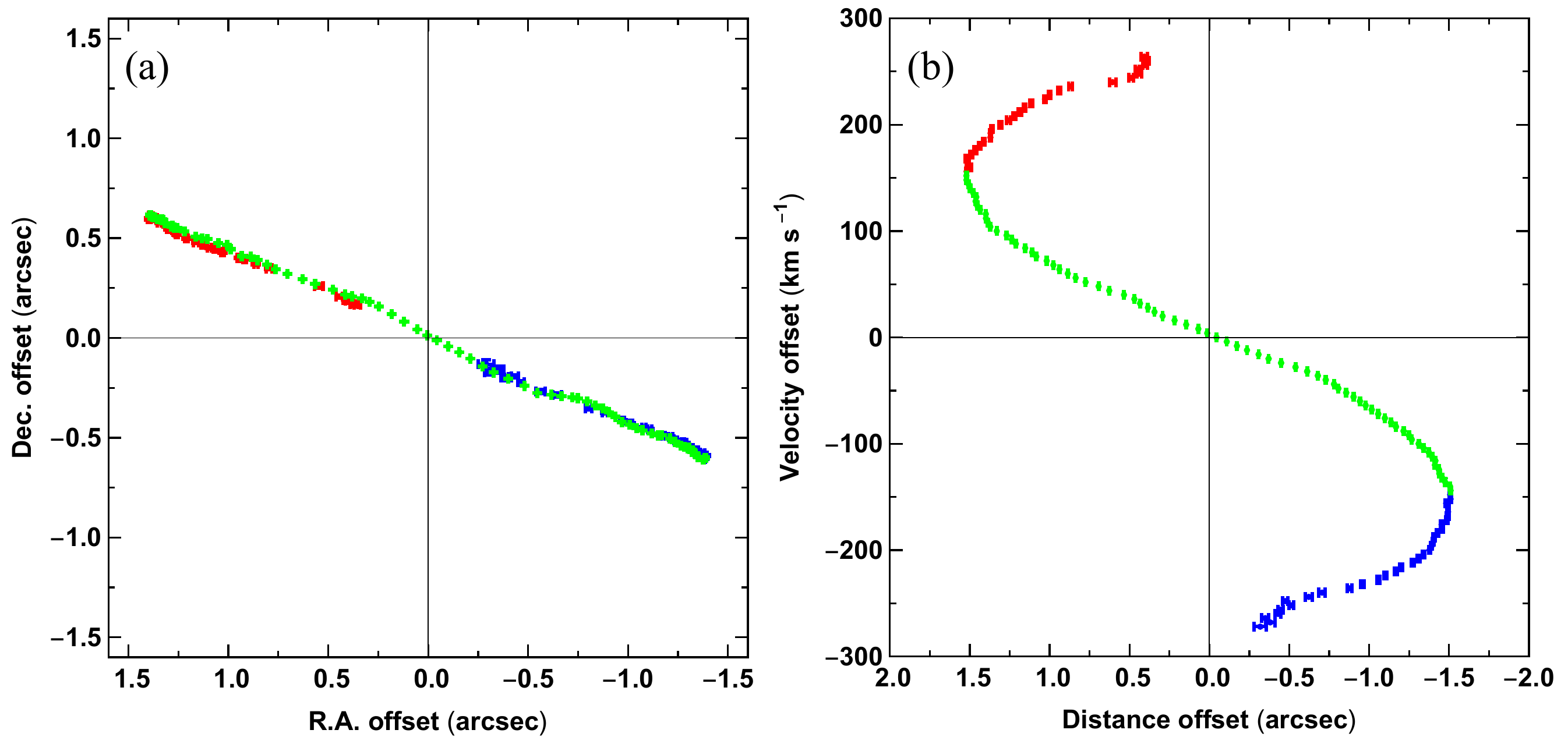}
	\caption{{\bf (a)} Peak positions of the brightness distribution of CO(3$\rightarrow$2) emission for 
		each individual velocity channel obtained with the task JMFIT of AIPS. The origin corresponds to the non-weighted 
		mean position of the distribution (see main text). The negative R.A. offsets are on the right side of the plot. 
		{\bf (b)} Distance offsets from the mean position as a function of 
		the velocity offset from the systemic velocity v$_{\rm sys, LSR}$=45~km~s$^{-1}$. The colour of the data points 
		are coded according to their velocity gradient, $|$dv$|$/$|$dr$|$; green for $|$dv$|$/$|$dr$|$>0 and blue-red
		for $|$dv$|$/$|$dr$|$<0. The negative distance 
		offsets correspond to data points with negative R.A. offsets and they are on the right side of the plot. The error 
		bars of the positions are included in the plot but their sizes is similar to the size of the data points. }
	\label{Fig1}%
\end{figure*}

The Hubble Space Telescope (HST) and Keck Adaptive Optics (AO) images of IRAS~16342$-$3814 show a small 
(3$^{\prime\prime}$) bipolar nebula, with the lobes separated by a dark equatorial waist \citep{Sahai1999,Sahai2005}. 
The morphology seen in the images implies that the lobes are bubble-like reflection nebulae illuminated by starlight 
escaping through polar holes in a dense, dusty waist obscuring the central star. The AO observations reveal a 
corkscrew-shaped pattern apparently etched into the lobe walls, which is inferred to be the signature of an underlying 
precessing jet. This jet has presumably carved out the observed bipolar cavities in the surrounding envelope 
formed during the AGB phase of the star \citep{Sahai2005}. \cite{Gledhill2012} studied the H$_{2}$ emission of this 
source and concluded that it is likely that the jet and outflow are powered by accretion onto a binary companion. The water 
masers in IRAS~16342$-$3814 are thought to be tracing bow shocks at the ends of a jet. From the analysis of the 
proper motions of the masers \cite{Claussen2009} derived an expansion velocity for the material in the head of the jet of 
v$_{\rm exp}$$\sim$155-180~km~s$^{-1}$. 

Recently, \cite{Sahai2017} carried out observations of the molecular emission of IRAS~16342$-$3814 with ALMA. From 
the results of their observations these authors concluded that the emission originates in a precessing molecular outflow 
with a wide precession angle of $\sim$90$^{\circ}$.  However, it is difficult to explain the creation of an outflow with 
such a wide precession angle by the jet underlying the corkscrew pattern seen from the AO observations. In this work 
we present a different spatio-kinematical model of the collimated molecular outflow in the water-fountain nebula 
IRAS~16342$-$3814 that considers a small opening angle, which is in better agreement with the results 
from previous observations.

\section{CO ALMA data of IRAS~16342$-$3814}

We retrieved archive ALMA data of the project 2012.1.00678.S (PI: R. Sahai) to study in more detail the kinematics of the 
molecular gas in IRAS~16342$-$3814. The spectral setup, sensitivity, angular resolution and results of the observations 
of the CO($J$=3$\rightarrow$2), $^{13}$CO($J$=3$\rightarrow$2) and H$^{13}$CO($J$=4$\rightarrow$3) emission lines were 
reported by \citep{Sahai2017}. In this work we further analyse the kinematics of the molecular gas using the emission from 
the CO($J$=3$\rightarrow$2) line, as this emission line has the highest signal-to-noise ratio. Since the position-velocity 
(P-V) diagrams of the $^{13}$CO($J$=3$\rightarrow$2) and H$^{13}$CO($J$=4$\rightarrow$3) are very similar to that of the 
CO($J$=3$\rightarrow$2) line, the main results presented in this work are valid for the three lines. 

In addition to the archival data, we have also obtained data from ALMA observations of the CO($J$=1$\rightarrow$0) line. 
These observations are part of the project 2018.1.00250.S (PI: D. Tafoya) and were carried out on December 9 2018 using 43-12m 
antennas with a maximum and minimum baselines of 740~m and 15~m, respectively. The resulting angular resolution 
was around 1$^{\prime\prime}$ and the maximum recoverable scale was 11$\rlap{.}^{\prime\prime}$8. The data were calibrated 
with the ALMA pipeline using the CASA version 5.4.0-68. Subsequently, the data were self-calibrated in phase following the 
standard self-calibration guidelines for ALMA data. In the present work, the observations of the CO($J$=1$\rightarrow$0) line
are used mainly to support the conclusions drawn from the CO($J$=3$\rightarrow$2) emission (see \S\ref{discussion_section}), 
but they will be presented and discussed in more detail somewhere else.  

\begin{figure*}
	\centering
	\includegraphics[angle=-0,scale=0.8]{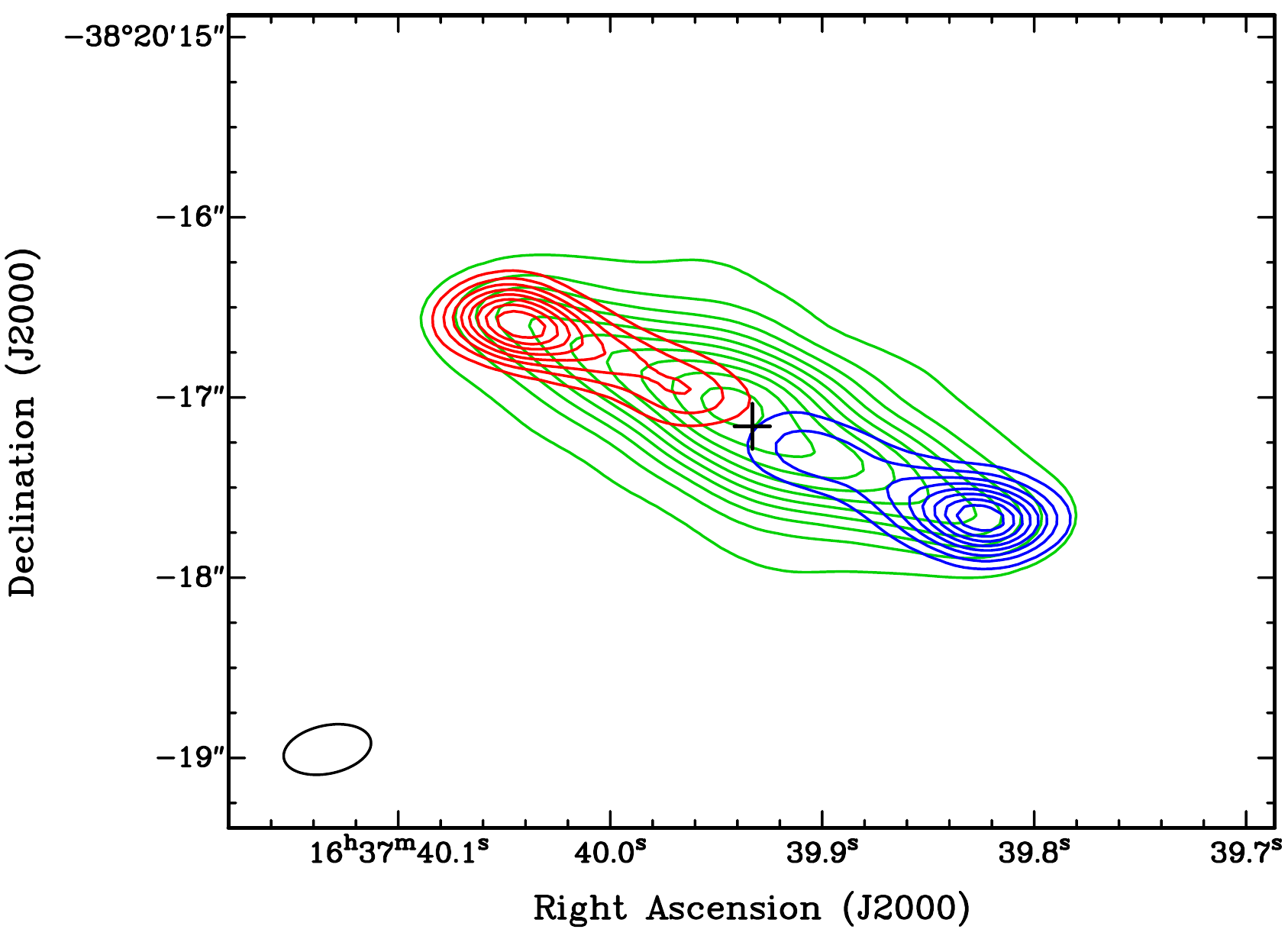}
	\caption{Velocity integrated CO($J$=3$\rightarrow$2) emission of IRAS~16342$-$3814. The blue, green and red contours 
		correspond to emission at velocity ranges $-$270<v$_{\rm offset}$(km~s$^{-1}$)<$-$150, 
		$-$150<v$_{\rm offset}$(km~s$^{-1}$)<+150;  and +150<v$_{\rm offset}$(km~s$^{-1}$)<+270, respectively 
		(see main text). The levels of the blue and red contours go from 20\% to 90\% the peak value of the emission (8.3 and 
		6.6~Jy~beam$^{-1}$~km~s$^{-1}$, respectively) at steps of 10\%. The levels of the green contours go from 5\% to 95\% 
		the peak value of the emission (50.7~Jy~beam$^{-1}$~km~s$^{-1}$) at steps of 10\%. The rms value of the 
		velocity-integrated maps is $\sim$7$\times$10$^{-2}$~Jy~beam$^{-1}$~km~s$^{-1}$. The cross indicates the peak of the 
		continuum emission. The ellipse located at the lower left corner indicates the synthesised beam, 
		$\theta_{\rm FWHM}$=0$\rlap{.}^{\prime\prime}$49$\times$0$\rlap{.}^{\prime\prime}$27, P.A.=$-$78$^{\circ}$}.
	\label{Fig2}%
\end{figure*}

From the data cube of the CO($J$=3$\rightarrow$2) line we extracted the peak position of the brightness distribution 
in every channel within the velocity range $-235$<v$_{\rm LSR}$(km~s$^{-1}$)<+325 using the task JMFIT of 
AIPS\footnote{AIPS is produced and maintained by the National Radio Astronomy Observatory, a facility of the National 
Science Foundation operated under cooperative agreement by Associated Universities, Inc.}. The emission within this 
velocity range corresponds to the high velocity outflow (HVO) observed by \cite{Sahai2017}, see their Figs. 1b and 2. 
It should be noted that apart from this emission, \cite{Sahai2017} also identified an extreme high-velocity 
outflow (EHVO) with higher expansion velocities. However, here and in the following sections we will only consider 
the emission associated to the HVO and leave the discussion of the nature of the EHVO for \S\ref{EHVO_section}. From the measured 
peak positions of the CO($J$=3$\rightarrow$2) emission we estimated the centroid of the distribution by computing 
the non-weighted mean value. The position of the centroid was used as a reference position throughout the analysis described 
below. In Fig.~\ref{Fig1}a we plot the declination (Dec.) and right ascension (R.A.) offsets of the emission peak 
positions with respect to the reference position. Subsequently, we calculated the separation of the emission peak 
positions to the reference position as  Dist.offset=$\pm$$\sqrt{{\rm R.A.offset}^{2}+{\rm Dec.offset}^{2}}$. We 
used the minus sign when R.A. offset was negative, otherwise we used the plus sign. In Fig.~\ref{Fig1}b we plot the 
distance-offset of the emission peak position as a function of the velocity-offset of the channel. The 
velocity-offsets are defined with respect to the systemic velocity of the source, assumed to be 
v$_{\rm sys, LSR}$=45~km~s$^{-1}$ \citep{Sahai2017}. The data points are colour-coded according to their
velocity gradient, $|$dv$|$/$|$d$r$$|$; green for $|$dv$|$/$|$d$r$$|$>0 and blue-red for $|$dv$|$/$|$d$r$$|$<0 
(see below). The distribution of the points in the plot of Fig.~\ref{Fig1}b exhibits an S-like shape similarly to the 
emission in the P-V diagram obtained by \cite{Sahai2017} \citep[see Fig. 2 of][]{Sahai2017}. This is due to the fact 
that the plot shown in Fig.~\ref{Fig1}b is equivalent to a P-V diagram since the  emission peak positions lie basically 
along one single direction on the sky (P.A. $\sim$67$^{\circ}$, see Fig.~\ref{Fig1}a).   

\begin{figure*}
	\centering
	\includegraphics[angle=0,scale=0.6]{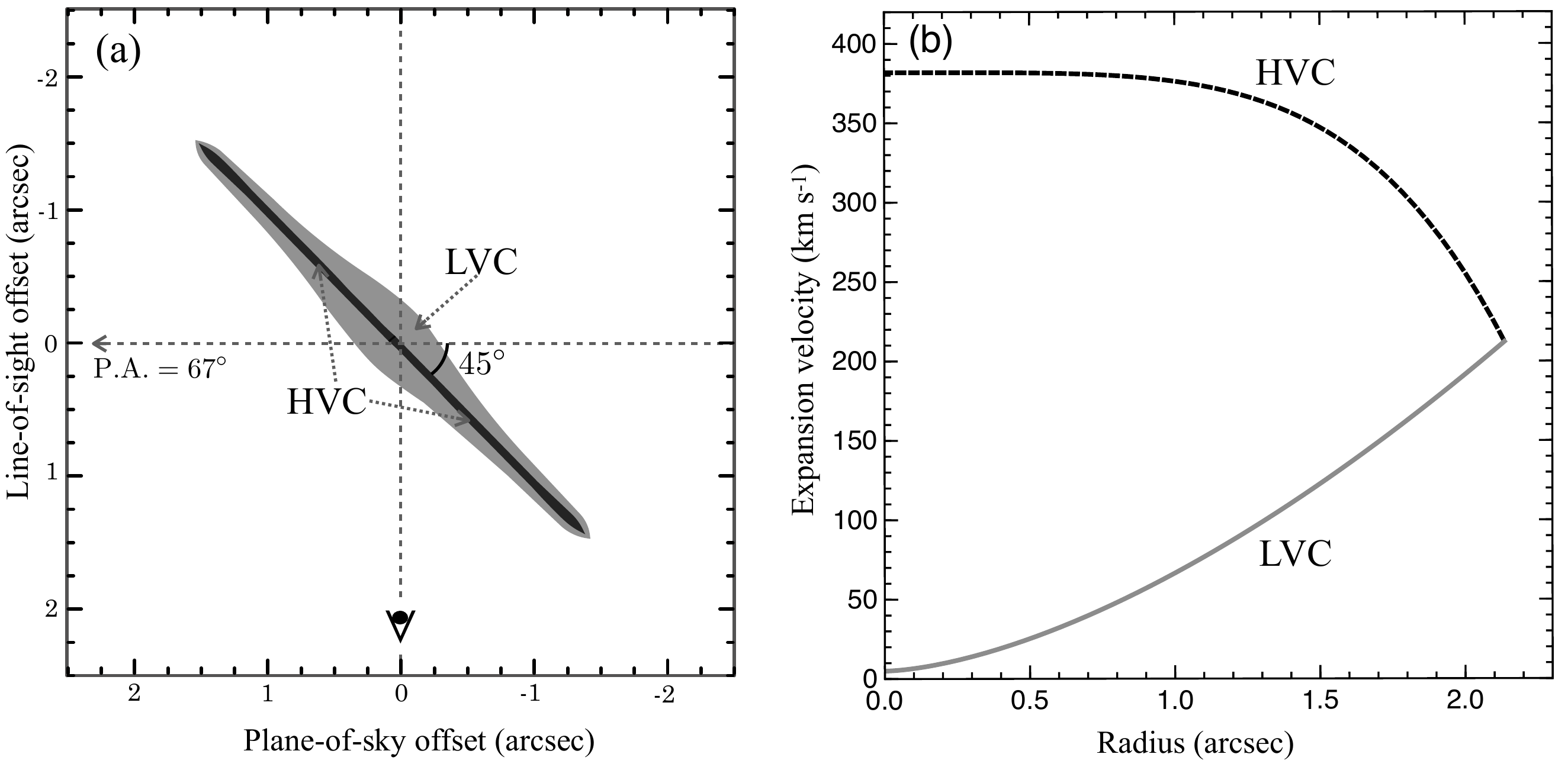}
	\caption{{\bf (a)} Geometry of the spatio-kinematical model of the molecular envelope of IRAS~16342$-$3814. 
		Note that the y-axis is angular size along the line-of-sight and the x-axis is angular size on the plane of the sky. 
		We assume an inclination of the major axis of the nebula of 45$^{\circ}$ with respect to the plane of the 
		sky \citep{Claussen2009}. The grey region indicates the LVC component. The black region indicates the HVC 
		component. {\bf (b)} Expansion velocity as a function of the separation from the center of the system. The grey 
		and black lines are the velocity laws for the LVC and HVC components, respectively. The mathematical expression 
		for the velocity laws is given in Equation~\ref{Eq:1}.}
	\label{Fig3}%
\end{figure*}

From the plot of Fig.~\ref{Fig1}b it can be seen that for velocities |v$_{\rm offset}$|<150~km~s$^{-1}$ 
the absolute value of the velocity offset increases monotonically as a function of the absolute distance offset, 
$|$dv$|$/$|$dr$|$>0. For |v$_{\rm offset}$|>150~km~s$^{-1}$ the gradient is opposite,  $|$dv$|$/$|$dr$|$<0. 
Therefore, this plot suggests the presence of different molecular gas components with different kinematical signatures, 
which are indicated with green and blue-red colours, respectively. We will refer to these components as kinematical 
components. In the following we consider these two types of kinematical components, one with positive velocity 
gradient and the other with negative velocity gradient. Surprisingly, as the plot of Fig.~\ref{Fig1}a shows, even 
though these components have different kinematical signatures, they are almost perfectly aligned on one single 
direction on the plane of the sky. 

In order to visualise the spatial distribution of the emission of the kinematical components defined above, we 
created velocity-integrated images of the CO($J$=3$\rightarrow$2) emission taking into account their corresponding velocity ranges. 
We considered three velocity ranges as follows: $-$150<v$_{\rm offset}$(km~s$^{-1}$)<+150; 
$-$270<v$_{\rm offset}$(km~s$^{-1}$)<$-$150 and +150<v$_{\rm offset}$(km~s$^{-1}$)<+270.  The first velocity range 
corresponds to the kinematical component with positive velocity gradient while the two last ones correspond 
to  the kinematical component with negative velocity gradient.

In Fig.~\ref{Fig2} we show contour maps of the integrated CO($J$=3$\rightarrow$2) emission obtained from the velocity 
ranges considered above. The blue contours indicate emission integrated over the velocity range 
$-$270<v$_{\rm offset}${\rm(km~s$^{-1}$)}<$-$150; the green contours indicate emission integrated over the 
velocity range $-$150<v$_{\rm offset}${\rm(km~s$^{-1}$)}<+150, and the red contours indicate emission integrated 
over the velocity range +150<v$_{\rm offset}${\rm(km~s$^{-1}$)}<+270. Note that this colour coding is the same as the 
one for the kinematical components shown in Fig.~\ref{Fig1}. From Fig.~\ref{Fig2} it can be seen that the 
emission with velocity-offsets |v$_{\rm offset}$|>150~km~s$^{-1}$, from now on referred to as high-velocity emission, 
traces a pair of collimated bipolar lobes along the direction with P.A.=$\sim$67$^{\circ}$. The velocity-integrated 
intensity of the high-velocity emission has two emission peaks located at the tips of the bipolar lobes. The geometric centre of 
the bipolar lobes coincides with the central part of the nebula, which is indicated with a cross that corresponds to 
the peak position of the continuum emission, (J2000) R.A.=16:37:39.935, Dec.=$-$38:20:17.15. The structure traced 
by emission with velocity offsets |v$_{\rm offset}$|<150~km~s$^{-1}$, from now on referred to as low-velocity emission, 
exhibits a more extended morphology encompassing the high-velocity emission. The low-velocity emission has only one 
intensity peak located near the centre of the system. The width of the brightness distribution of the low-velocity emission 
decreases along the direction of the bipolar axis resembling an elongated diamond-like shape.
  
\section{Spatio-kinematical model of the molecular gas}\label{spt_kin_model_section}

In this section we present a simple spatio-kinematical model of the molecular gas of IRAS~16342$-$3814 to 
reproduce the spatial distribution and velocity gradients seen in Figs.~\ref{Fig1} and \ref{Fig2}. In the following 
we describe the considerations that we made to define the morphology and velocity field used in our model. 
First we note from Fig. ~\ref{Fig1}a  that the spatial distribution of the emission peak positions of the kinematical 
components, defined in \S2, lie along 
one single direction on the plane of the sky. This could be due to the kinematical components being intrinsically 
aligned on one single direction, or alternatively they could lie on different directions but appear perfectly aligned 
on the plane of the sky. We consider that the latter possibility is very unlikely as even a small inclination of the system 
with respect to the line-of-sight would result in the components appearing misaligned on the plane of the sky. 
Furthermore, previous observations suggest that the gas and dust are located within a narrow cone elongated along 
the direction with P.A.$\sim$67$^{\circ}$ 
\citep{Sahai2005,Claussen2009,Imai2012, Gledhill2012}. Therefore, in this work we adopt a geometry similar to the 
one assumed in previous works and propose that the observed CO($J$=3$\rightarrow$2) emission can be explained 
using two morphological structures associated to the kinematical components defined in \S2: i) an elongated structure 
with a relatively wide waist and narrow tips. This component is associated with the low-velocity emission and we will 
refer to it as the low-velocity component (LVC). ii) A pencil-like collimated structure associated with the high-velocity 
emission that lies embedded within the LVC component. We will refer to this component as the high-velocity 
component (HVC). Thus, the morphology for the CO emitting region of IRAS~16342$-$3814 in our spatio-kinematical 
model has a shape that resembles a French spindle, as it is shown in Fig.~\ref{Fig3}a. 
 
The width of the LVC, as a function of the velocity, was determined by fitting a two-dimensional Gaussian function to 
the intensity distribution in each individual channel of the data cube. The size was set equal to the quadratic average of 
the major and minor axis, deconvolved from the beam, of the fitted elliptical Gaussian function. The width of this component, 
at its waist, is $\sim$0$\rlap{.}^{\prime\prime}$5 and it decreases monotonically to reach a value of $\sim$0$\rlap{.}^{\prime\prime}$15 
at the polar tips. The width of the HVC is not resolved by these observations but we adopted a width at the equator of 
$\sim$0$\rlap{.}^{\prime\prime}$02 and set it to increase monotonically to a value of $\sim$0$\rlap{.}^{\prime\prime}$05 based on 
the opening angle of the sinusoidal distribution of the peak positions of this component (see \S4.2 and Fig.~\ref{Fig8}). The major axis 
of both components was assumed to have an inclination angle with respect to the plane of the sky $\theta_{\rm inc}$=45$^{\circ}$, 
which \cite{Claussen2009} obtained by measuring the inclination angle of the three-dimensional velocity vectors of the H$_{2}$O 
masers. It is worth mentioning that an independent estimate of the inclination angle of  $\sim$40$^{\circ}$ was obtained by \cite{Sahai1999} 
from observations of OH masers. However, we opted for the value derived by \cite{Claussen2009} 
because it was obtained in a more straightforward manner. The de-projected length along the major axis of both components was 
estimated from the maximum separation of the emission peak positions from the reference position (see Fig.~\ref{Fig1}b), 
$r_{\rm max}$=1$\rlap{.}^{\prime\prime}$5/cos\,$\theta_{\rm inc}$=2$\rlap{.}^{\prime\prime}$12. In this work we adopt a distance 
to IRAS~16342$-$3814 $D$=2~kpc \citep{Sahai1999}. To express the expansion velocity field of the molecular gas we used 
a power-law function given by the following 
equation: 

\begin{equation}\label{Eq:1}
\left[\frac{{\rm v}_{\rm exp}(r)}{{\rm km~s}^{-1}}\right]={\rm v}_{\rm i}-({\rm v}_{\rm i}-{\rm v}_{\rm f})\times\left(\frac{r}{r_{\rm max}}\right)^{\eta}, 
\end{equation}
where v$_{\rm i}$ (initial velocity) is the value of the expansion velocity at $r$=0; v$_{\rm f}$ (final velocity) 
is the value of the velocity at the tips of the lobes, and $\eta$ is the exponent of the power-law function.

\begin{figure*}
	\centering
	\includegraphics[angle=0,scale=0.61]{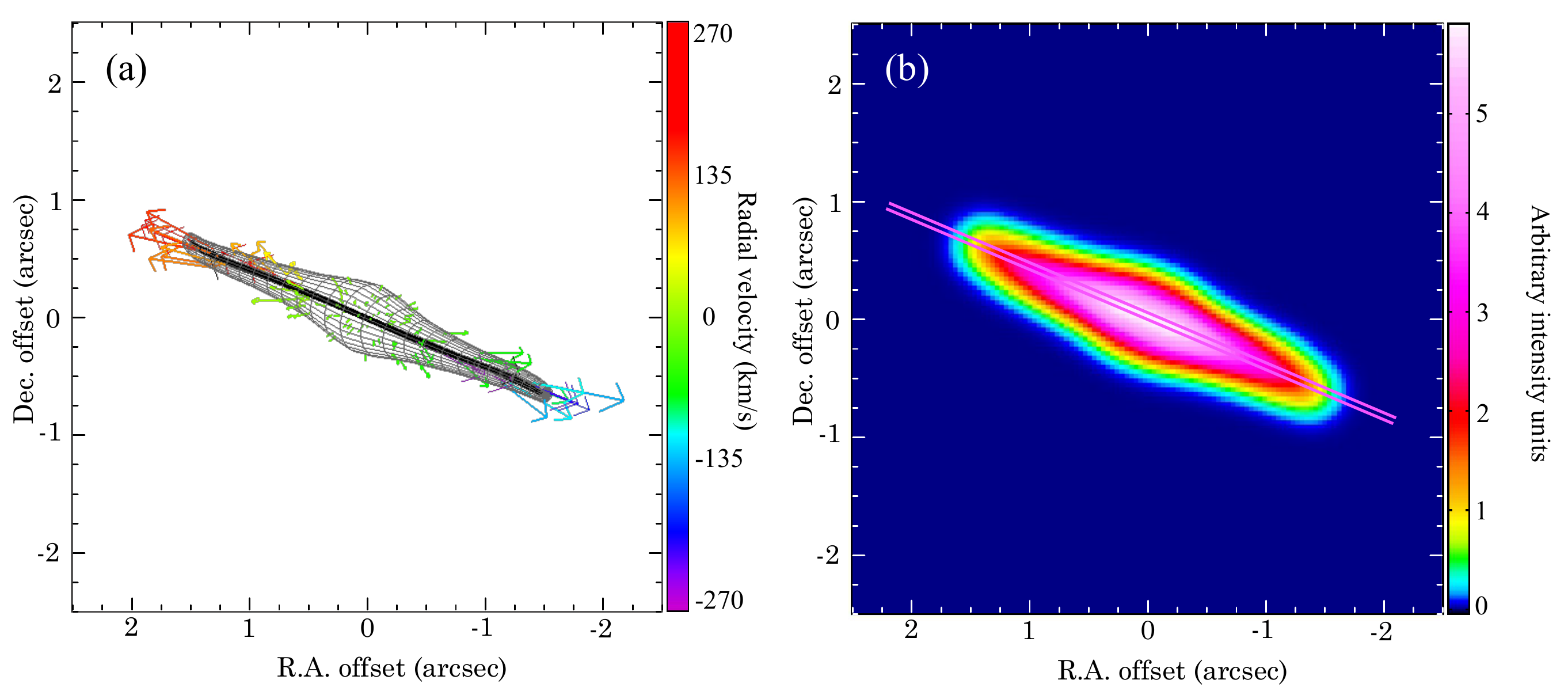}
	\caption{{\bf (a)} Mesh representation of the spatial distribution of the molecular outflow of IRAS~16342$-$3814. 
	The HVC and LVC are shown in black and grey colours, respectively. The vectors represent the expansion velocity 
	of the gas and they are coloured according to their component in the line-of-sight. {\bf (b)} Brightness distribution 
	of the CO emission obtained from the render function of SHAPE. The intensity is given in arbitrary units. The synthetic 
	image was convolved with a Gaussian function with FWHM similar to the synthesised beam of the observations. 
	The parallel lines at P.A.=67$^{\circ}$ indicate the direction of a slit used to generate the synthetic P-V diagram 
	shown in Fig.~\ref{Fig5}a.}
	\label{Fig4}%
\end{figure*}
\begin{figure*}
	\centering
	\includegraphics[angle=0,scale=0.6]{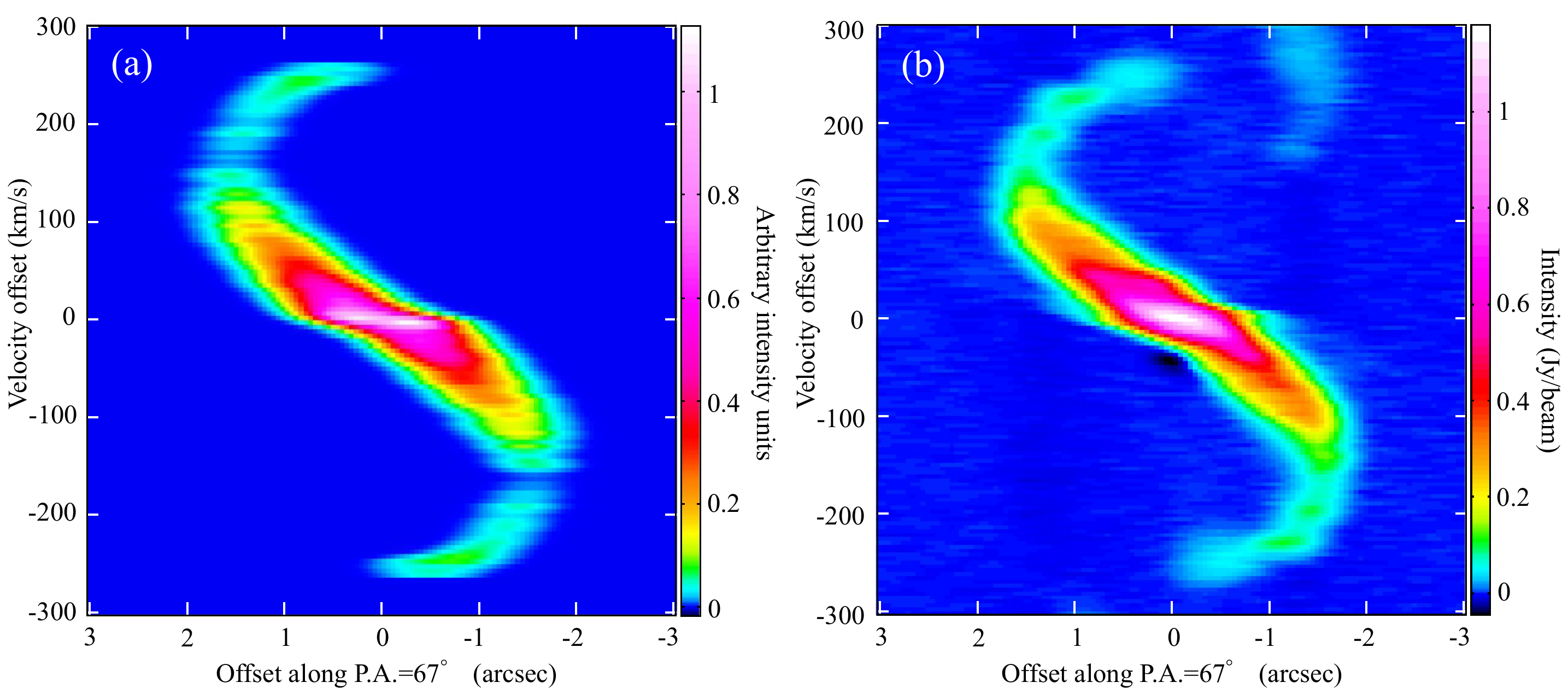}
	\caption{{\bf (a)} P-V diagram of IRAS~16342$-$3814 obtained from our SHAPE model. {\bf (b)} 
		P-V diagram of IRAS~16342$-$3814 from the ALMA observations. The faint emission in 
		the upper-right corner is due to the H$^{13}$CO($J$=4$\rightarrow$3) line. 
		Both P-V diagrams were obtained using a slit along P.A.=67$^{\circ}$}
	\label{Fig5}
\end{figure*}

The values of the initial and final velocity for the HVC can be obtained from the maximum and minimum value 
of the velocity range of its corresponding kinematical component, i.e. 150<|v$_{\rm offset}$ (km~s$^{-1}$)|<270~(see \S2). 
Thus, taking into account the inclination angle, v$_{\rm i, HVC}$=270~km~s$^{-1}$/sin\,$\theta_{\rm inc}$=382~km~s$^{-1}$ 
and v$_{\rm f, HVC}$=150~km~s$^{-1}$/sin\,$\theta_{\rm inc}$=212~km~s$^{-1}$. Given that the data points in 
Fig.~\ref{Fig1}b exhibit a continuous transition between the two kinematical components, we set the final velocity of the 
LVC equal the the final velocity of the HVC, i.e. v$_{\rm f}$$\equiv$v$_{\rm f, LVC}$=v$_{\rm f, HVC}$. The initial 
velocity for the LVC, ${\rm v}_{\rm i, LVC}$, cannot be directly obtained from the observations. From the plot of 
Fig.~\ref{Fig1}b it can be seen that ${\rm v}_{\rm exp}$($r$=0)$\approx$0, but the points in this plot represent an average 
position of the distribution of the emission for a certain velocity offset. 
Therefore, the point located at the origin of coordinates, (v=0, $r$=0), does not necessarily indicate that 
v$_{\rm exp}$=0 when $r$=0, but it means that the average position of the emission is equal to zero in the channel with 
v$_{\rm offset}$=0. Consequently, v$_{\rm i, LVC}$ was left as a free parameter, together with the exponents of the 
power-law functions, to be adjusted by further comparison of our model with the observations.

\subsection{3D modelling with SHAPE}

We used the spatio-kinematic modelling tool SHAPE \citep{Steffen2011} to construct a three-dimensional (3D) version 
of our model of the distribution and kinematics of the molecular gas in IRAS~16342$-$3814 and compare it with the
observations. It should be remarked 
that this model is not meant to reproduce the emission of any particular molecular line but it only illustrates the 
overall behaviour of the radiation arising within the kinematical components. First we 
obtained synthetic images and P-V diagrams of our spatio-kinematical model with SHAPE using initial guess values. Subsequently, we compared 
the results of SHAPE with the observations in an iterative approach to derive the values of the free parameters of our model 
(see below). The synthetic P-V diagram obtained with SHAPE and the one from the observations were compared by eye. 
After several iterations we found a good fit to the observations using the following parameters for the power-law function: 
v$_{\rm i, HVC}$=382~km~s$^{-1}$, v$_{\rm f, HVC}$=212~km~s$^{-1}$ and $\eta_{\rm HVC}$=4.5 for the 
HVC; v$_{\rm i, LVC}$=5~km~s$^{-1}$, v$_{\rm f, LVC}$=212~km~s$^{-1}$ and $\eta_{\rm LVC}$=1.6 for the LVC. 
The power-law functions for both components are shown in the plot of Fig.~\ref{Fig3}b.

Fig.~\ref{Fig4}a shows a mesh view of the 3D morphology of the emitting region used in SHAPE. The grey and black meshes
represents the LVC and HVC components, respectively. The arrows represent vectors of the velocity field of the 
expanding gas. The colours of the velocity vectors are coded according to their component in the line-of-sight 
direction. We performed a simple radiative transfer of our model using the physical render function of SHAPE, thus 
the emerging emission is proportional to $T_{\rm ex} (1-e^{-\tau_{\nu}})$, where $T_{\rm ex}$ is the excitation temperature 
and $\tau_{\nu}$ is the optical depth. The LTE option in SHAPE was turned on, which calculates the absorption/emission ratio 
assuming Local Thermodynamic Equilibrium conditions, and the absorption coefficient was defined so 
that $\tau_{\nu}$ is proportional to the product of the gas density and the length of the emitting region along the 
line-of-sigh, $\tau_{\nu} \propto n$$\times$$l$. For simplicity, the density of the gas and the excitation temperature 
of the HVC were assumed to be constant throughout the outflow. At the base of the LVC, the excitation temperature 
was set to a value 5 times lower than the HVC, increasing monotonically as a function of the distance to reach the 
same value as the HVC at the tip of the lobes. We aware the reader that it is not the scope of this work to perform 
a detailed radiative transfer analysis of the model but we are only concerned about the kinematics of the gas. 
Therefore, the resulting intensity of the synthetic images is given in arbitrary units and do not necessarily correspond 
to the intensity of the observed emission. The synthetic images were convolved with a Gaussian function with FWHM 
similar to the synthesised beam of the observations. 

\begin{figure*}
	\centering
	\includegraphics[angle=0,scale=0.9]{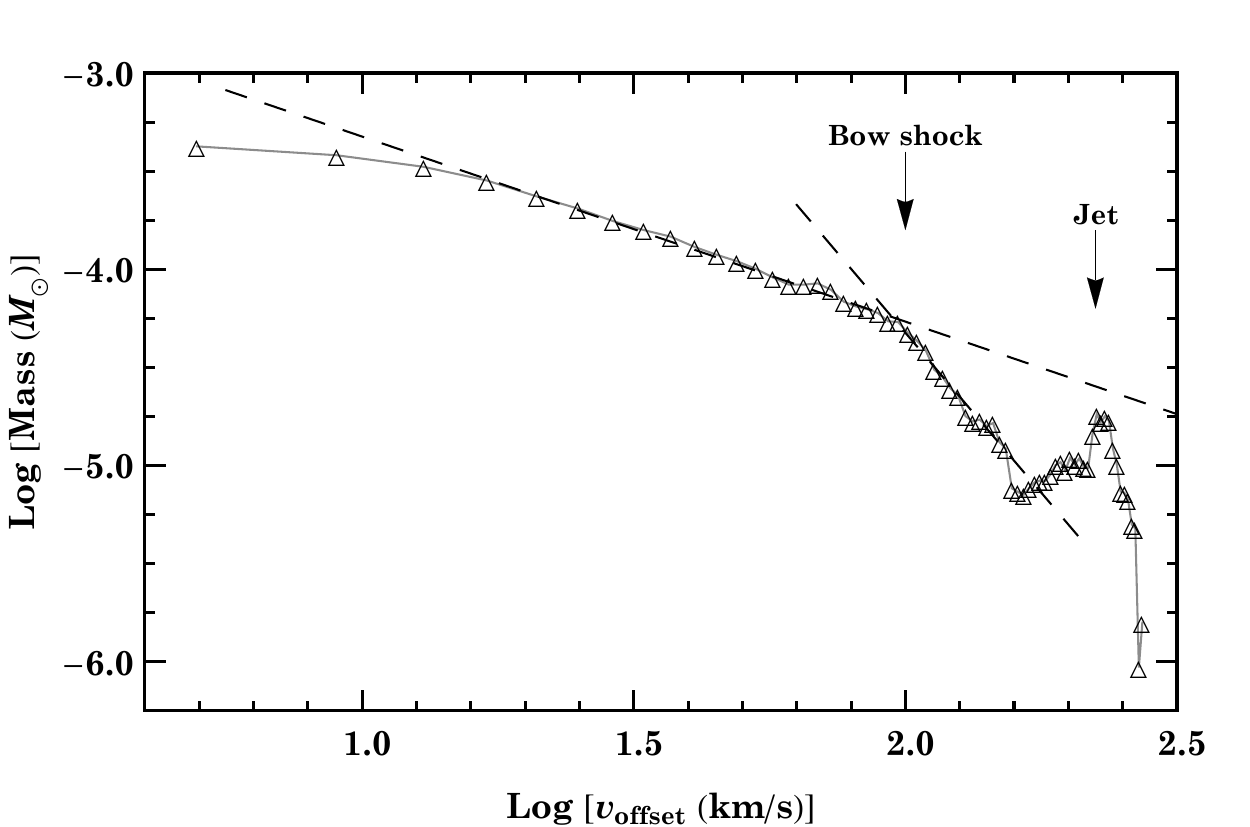}
	\caption{Molecular mass as a function of expansion velocity obtained from the emission blue-shifted with respect to the 
		systemic velocity. The dashed lines are power-law fits to the data in the velocity ranges 
		10<v$_{\rm offset}$(km~s$^{-1}$)<100 and 100<v$_{\rm offset}$(km~s$^{-1}$)<150, respectively. The arrows indicate 
		the emission associated with the bow shock and the jet, respectively (see main text).}
	\label{Fig6}%
\end{figure*}

In Fig.~\ref{Fig4}b we show a synthetic image of the emission of the source on the plane of the sky. The spatial 
distribution of the emission of our model resembles the image of the HVO of IRAS~16342$-$3814 obtained by 
\cite{Sahai2017} \cite[see Fig. 2b of][]{Sahai2017}. The two parallel lines in Fig.~\ref{Fig4}b indicate the direction of a slit used 
to generate the synthetic P-V diagram shown in Fig.~\ref{Fig5}a. We used a similar slit to generate a P-V diagram 
from the data cube of the CO($J$=3$\rightarrow$2) emission, which is shown in Fig.~\ref{Fig5}b. The faint emission in 
the upper-right corner of the P-V diagram of Fig.~\ref{Fig5}b is due to the H$^{13}$CO($J$=4$\rightarrow$3) line 
overlapping in frequency with the CO($J$=3$\rightarrow$2) emission \citep[see also Figure 3 of][]{Sahai2017}. The synthetic P-V diagram in 
Fig.~\ref{Fig5}a reproduces very well the main features of the CO($J$=3$\rightarrow$2) emission P-V diagram. The 
clumpy structure that some parts of the  synthetic P-V diagram exhibit is due to the binning of the velocity 
channels during the convolution of the model.

\section{Discussion and interpretation of the spatio-kinematical model}\label{discussion_section}

\cite{Sahai2017} proposed a model to explain the P-V diagrams of the CO($J$=3$\rightarrow$2) and H$^{13}$CO($J$=4$\rightarrow$3) 
emission assuming that all the molecular gas is expanding with the same intrinsic velocity. In their model, the S-like pattern seen in 
the P-V diagram results from the projection effect of the material moving along different directions with respect to the line-of-sight. 
Under this assumption they concluded that the morphology of the source is a S-like bipolar shape with some gas 
clouds expanding in directions near the line-of-sight and some other ones near the plane of the sky 
\citep[see Fig. 3 of][]{Sahai2017}. This model has two main problems: i) The constant expansion velocity assumption is 
most likely not true, since the velocity of the gas would depend on the hydrodynamic interaction of the driving jet and 
the CSE, as \cite{Sahai2017} acknowledge. ii) The morphology of the nebula derived from this assumption may result 
in an S-like shape with a very wide opening angle ($\sim$90$^{\circ}$) that would not be consistent with the highly 
collimated morphologies suggested by the IR observations and the H$_{2}$O masers. However, we note that a possible 
resolution to the problem in item (ii) is provided by the Spiral-class PNe introduced by 
\cite{Sahai2011} \citep[e.g. PK 032+07\#2 and PNG 356.8+0.3.3, see Fig. 7 of][]{Sahai2011}, which, if viewed at a 
preferred orientation angle, could resemble the lobes seen in IRAS~16342$-$3814.

In the previous section, \S\ref{spt_kin_model_section}, we showed that the spatio-kinematical model presented in this work reproduces very 
well the distribution of the observed CO($J$=3$\rightarrow$2) emission in the P-V diagram. As discussed above, the gradients 
seen in the P-V diagram (Fig.~\ref{Fig5}b) are interpreted as being due to actual variations of the velocity field of the gas 
within an elongated narrow region. Therefore, our model circumvents the problem of assuming a constant expansion 
velocity and reconciles the morphology of the CO gas with the morphology of the source derived from other observations. 
In the following we discuss on the astrophysical origin of the observed velocity gradients. 
In the past three decades a wealth of observations toward regions of star formation, as well as toward evolved stars, 
have revealed that the main properties that characterise collimated molecular outflows are: i) The velocity field of the 
outflow is described by a proportional relationship between the expansion velocity and the distance to the driving 
source, i.e. v$\propto$$r^{\eta}$, with $\eta$$\approx$1. ii) The collimation of the flow increases systematically with 
flow velocity and distance from the driving source. iii) The flow exhibits a power-law variation of mass with velocity, 
i.e. $m(\rm v)$$\propto$v$^{-\gamma}$ \citep[][and references therein]{Lada1996, Cabrit1997}. The most popular 
mechanism to explain these properties is the jet-driven bow shock \citep{Raga1993,Smith1997}. This mechanism 
naturally creates a positive velocity gradient since as one moves from the broader wings to the narrow apex of the bow 
shock the shock becomes less oblique and the net forward velocity increases. In addition, this mechanism produces 
more swept-up mass at low velocities since the intercepted mass flux is determined by the bow shock cross-section, 
which steadily grows in the bow wings while the velocity decreases \citep{Cabrit1997}. The LVC of IRAS~16342$-$3814 
exhibits all the properties that characterise jet-driven bow shocks. As shown in \S\ref{spt_kin_model_section}, the velocity 
field of the LVC has a power-law dependence with the distance, although the power-law exponent, $\eta$=1.6, is 
somewhat larger than the typical values seen in other molecular outflows. The collimation of the LVC increases with 
velocity as it can be seen in Fig.~\ref{Fig2}. Finally, assuming optically thin emission, LTE conditions and an average excitation 
temperature $T_{\rm ex}$=30~K, 
the mass of gas expanding at a given velocity, i.e. the mass spectrum $m$(v), can be obtained from the line profile 
of the CO($J$=3$\rightarrow$2) emission\footnote{The value of the mass estimated in this manner would represent a 
lower limit of the actual value if the emission is optically thick.}. In Fig.~\ref{Fig6} we show a log-log plot of the mass as 
a function of the velocity offset, where we have used a fractional abundance of CO relative to 
H$_{2}$, $f$(CO)=3$\times$10$^{-4}$ \citep{Sahai2017}. 
This plot reveals that the emission between 10<v$_{\rm offset}$(km~s$^{-1}$)<100~km~s$^{-1}$ is described by a 
power-law $m$(v)$\propto$v$^{-\gamma}$ with $\gamma$=0.94, and between 
100<v$_{\rm offset}$(km~s$^{-1}$)<150~km~s$^{-1}$ $\gamma$=3.2. This double power-law variation of the mass 
with velocity is exactly what observations have revealed in other molecular outflows 
\citep{Kuiper1981, Rodriguez1982,Lada1996}, and confirmed by numerical simulations of jet-driven molecular 
outflows \citep{Smith1997}. However, it should be pointed out that, if the CO($J$=3$\rightarrow$2) line is optically thick, 
the values of these power-law exponents represent just a lower limit. Furthermore, it cannot be ruled out that 
the break of the power-law exponent around v$_{\rm offset}$=100~km~s$^{-1}$ might be due to a change in the optical 
depth regime.

\begin{figure*}
	\centering
	\includegraphics[angle=0,scale=0.6]{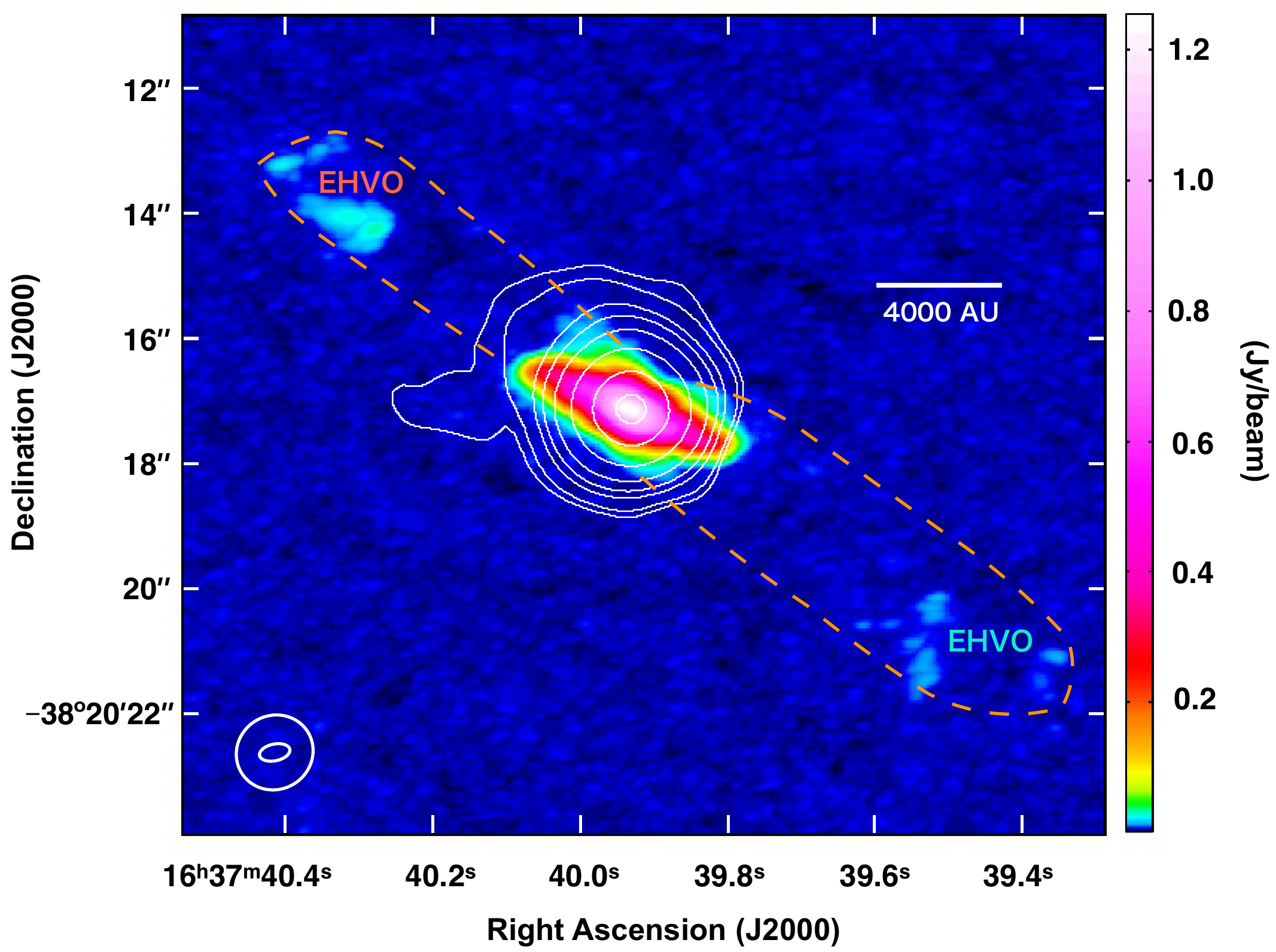}
	\caption{CO emission from IRAS~16342$-$3814. The colour scale corresponds to the moment-8 (maximum value of 
		the spectrum for each pixel) image of the CO($J$=3$\rightarrow$2) emission. The contours indicate the 
		CO($J$=1$\rightarrow$0) emission averaged over the velocity range $-$10<v$_{\rm offset}$(km~s$^{-1}$)<+10 with values 
		(0.007, 0.01, 0.02, 0.03, 0.05, 0.1, 0.2, 0.3) 
		Jy~beam$^{-1}$, where the first contour represents 4 times the value of the rms. The orange dashed line delineates 
		the region that contains emission from the EHVO. The ellipses located at the lower left corner indicate the synthesised beams 
		of the CO($J$=3$\rightarrow$2)  and CO($J$=1$\rightarrow$0) observations, respectively. The beam size of the 
		CO($J$=3$\rightarrow$2) observations is the same as in Fig.~\ref{Fig2} and the beam size of the CO($J$=1$\rightarrow$0) 
		observations is 1$\rlap{.}^{\prime\prime}$25$\times$1$\rlap{.}^{\prime\prime}$18 with P.A.=$-$58$^{\circ}$.}
	\label{Fig7}%
\end{figure*}

In Fig.~\ref{Fig7} we show contours indicating the CO($J$=1$\rightarrow$0) emission superimposed on a 
moment-8 (maximum value of the spectrum for each pixel) image of the CO($J$=3$\rightarrow$2) emission. The 
CO($J$=1$\rightarrow$0) emission was averaged over the velocity range $-$10<v$_{\rm offset}$(km~s$^{-1}$)<+10, 
which is the typical expansion velocity of the CSE formed during the AGB phase. The brightness distribution of the 
emission is round and has a deconvolved size of $\sim$0$\rlap{.}^{\prime\prime}$9, which is almost twice as large as the deconvolved 
size of the CO($J$=3$\rightarrow$2) emission in the same velocity range (see \S3), suggesting that it is tracing material of the CSE 
created by the AGB wind. The existence of a relic AGB circumstellar envelope is also suggested by the low-velocity 
OH~1612 maser emission, which exhibits maser features distributed over a region more extended than the bipolar outflow 
\cite[see Fig. 1 of ][]{Sahai1999}. Moreover, \cite{Murakawa2012} showed that the SED of IRAS~16342$-$3814 
is best reproduced by a model that includes a disk, a torus, bipolar lobes, as well as a spherical AGB envelope. 
Thus, given these arguments, together with the results presented in the previous paragraph, we conclude that the LVC in 
IRAS~16342$-$3814 corresponds to material of the CSE that has been swept-up by the jet-driven bow shock.

A feature revealed in these ALMA observations, not commonly found so evident in other molecular outflows, is the 
emission associated to the HVC. Fig.~\ref{Fig6} shows that for velocities offsets higher than 150~km~s$^{-1}$ 
the mass spectrum exhibits a plateau followed by a sharp peak. \cite{Smith1997} obtained a similar feature in the 
intensity profile from their hydrodynamic simulations of a jet-driven outflow. They identified this feature with the 
jet's direct emission \citep[see Fig. 1 and 2 of][]{Smith1997}. The deceleration of these component suggests the 
presence of turbulent mixing through Kelvin-Helmholtz instabilities at the velocity shear between the jet and the 
ambient gas. If turbulent entrainment along the beam is the dominant process, then the jet decreases in average 
velocity along its length as it transfers momentum to the entrained material \citep{Chernin1994}. Given a velocity 
of the jet v$_{\rm exp}$=380~km~s$^{-1}$, a sound speed in the medium $c_{s}$$\sim$1~km~s$^{-1}$, a 
density of the medium $n_{\rm H_{2}}$=10$^{6}$~cm$^{-3}$ and a kinetic temperature $T_{\rm k}$$\sim$100~K, 
the Reynolds number of the jet is $\mathcal{R}e_{m}$>10$^{10}$. Consequently, the jet is expected to be 
highly turbulent and suffer deceleration, which is indeed what the observations reveal. We conclude that the 
HVC represents material entrained by the underlying jet near its axis or it could be the molecular component of 
the jet itself.

As mentioned earlier, molecular outflows with the characteristic properties of a jet-driven outflow have also been 
found in the CSE of some evolved stars \citep[e.g.][]{Bujarrabal1998,Sanchez-Contreras2000,Alcolea2001, Alcolea2007}.
For these objects, the linear relationship between the expansion velocity and the distance to the driving source has 
been interpreted in terms of free expansion of the gas after having suffered a strong axial acceleration by 
a collimated wind or jet during a time period much shorter than the whole post-AGB lifetime of the source 
\citep{Bujarrabal1998,Alcolea2001}. Particularly, \cite{Alcolea2001} derived an upper limit to the duration of 
this post-AGB interaction in the pre-PN OH~231.8$+$4.2 of $\sim$125~years. Our results suggest that in 
IRAS~16342$-$3814 we are witnessing the presence of two different processes of entrainment, one called 
``prompt entrainment", which transfers momentum through the leading bow shock, and the second is 
``steady-state entrainment", which refers to ambient gas that is entrained along the sides of the jet 
\citep{De-Young1986,Chernin1994}. The former would lead to the formation of the linear relationship of the velocity 
with the distance seen in the PV diagrams of the molecular emission of the more evolved objects, which corresponds 
to swept-up material in a jet-driven bow shock. The new ingredient revealed in these ALMA observations is the HVC, 
which is related to the driving jet. It is likely that in the molecular outflows seen in other more evolved objects the 
jet has long disappeared as it is expected to last only for a couple of hundred years. This strengthens the idea that 
wf-nebula are indeed undergoing the ephemeral transition where they develop the collimated outflows that shape 
the CSE leading to the formation of asymmetrical PNe. In fact, recently Orosz et al. (2018) showed a beautiful example 
of a decelerating outflow in the wf-nebula IRAS~18113$-$2503 traced by H$_{2}$O masers, which is likely to be 
directly related to the driving jet.

\begin{figure*}
	\centering
	\includegraphics[angle=0,scale=0.9]{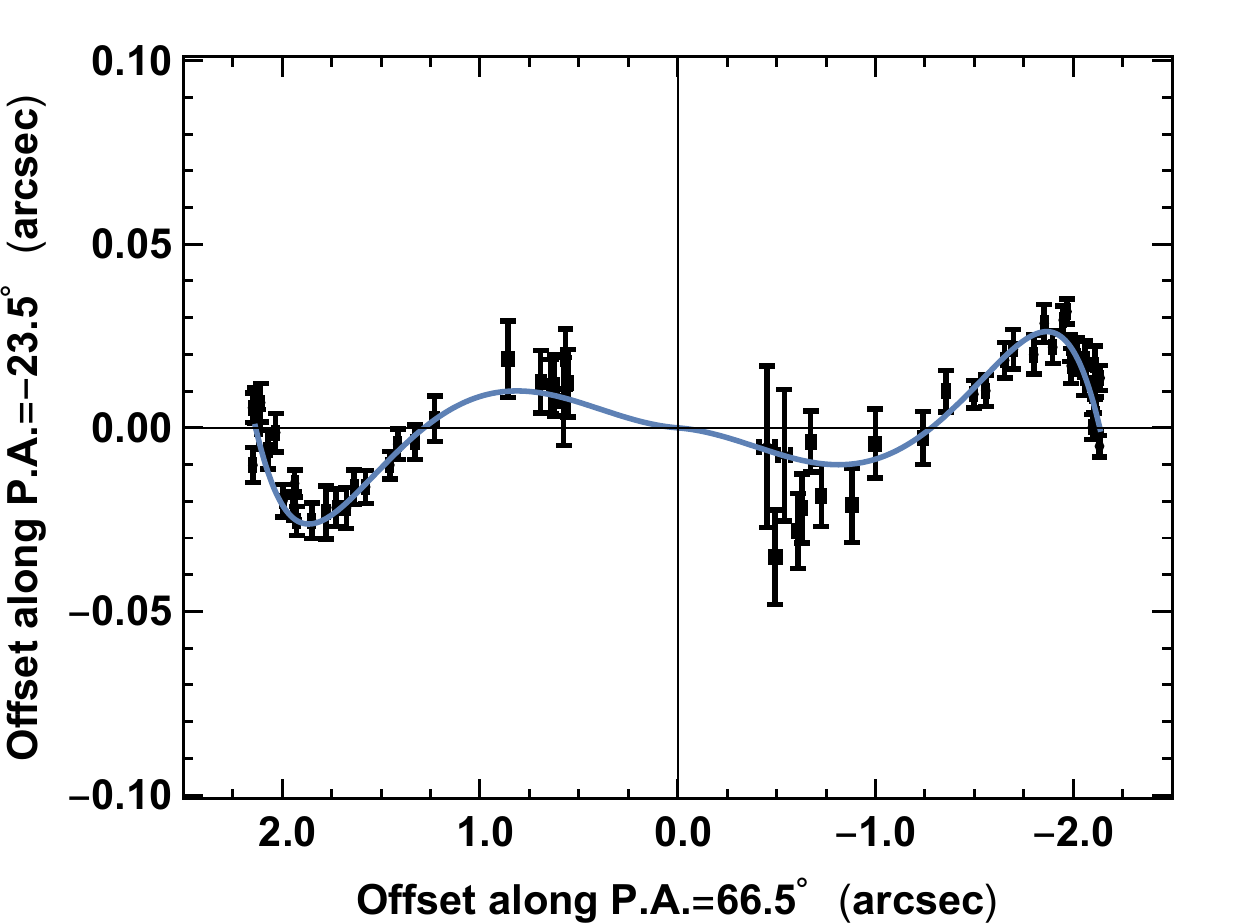}
	\caption{Oscillating pattern of the positions of the peak-emission for the high-velocity emission. The blue line indicates a 
		sinusoidal function fit to the data using Equation~\ref{Eq:2}.}
	\label{Fig8}%
\end{figure*}

\subsection{Timescale of the molecular jet and the equatorial waist}

The plot on Fig.~\ref{Fig3}b and the P-V diagram in Fig.~\ref{Fig5}b show an instantaneous picture of the 
velocity field in the outflow at present time. Thus they cannot be used to trace the velocity history of the individual gas 
particles, unless the deceleration law as a function of time is known. In order to understand the evolution of the P-V diagram 
as a function of time an entire knowledge of the complex dynamical system of the outflow is required. 
Consequently, it is difficult to estimate the exact lifetime of the outflow. However, if we make the reasonable assumption 
that the jet has been launched at roughly the same speed throughout its lifetime, it is clear that the gas particles located 
at the head of the jet suffered a deceleration from the initial velocity, v$_{\rm i}$, to the final velocity, v$_{\rm f}$, 
across its length, $r_{\rm max}$. If we consider a constant deceleration function, the lifetime of the outflow can be 
obtained as $\tau_{\rm jet}$=2$\times$$r_{\rm max}/({\rm v}_{\rm i}+{\rm v}_{\rm f})$$\approx$70~years.
This value is just a lower limit for the lifetime as it is likely that the jet suffered a steeper deceleration in a region closer 
to the star, where the density of the CSE is larger. For example, if the head of the jet decelerated to its current value, 
v$_{\rm f}$, soon after it was launched and it has been expanding at nearly constant velocity afterwards, the lifetime 
of the jet is $\tau_{\rm jet}$=$r_{\rm max}/{\rm v}_{\rm f}$$\approx$100~years.

\cite{Sahai2017} identified a toroidal structure in IRAS~16342$-$3814 from the emission of the 
H$^{13}$CO(4$\rightarrow$3) line and derived a radius $r_{\rm torus}$$\approx$0$\rlap{.}^{\prime\prime}$33 and an 
expansion velocity of v$_{\rm torus}$=20~km~s$^{-1}$, which gives an expansion timescale of the torus 
$\tau_{\rm torus}$$\approx$160~years. This value is larger than the timescale of the jet derived above. 
However, this timescale is obtained assuming that the gas of the torus expanded at constant velocity. According to 
our interpretation of the LVC (see \S\ref{discussion_section}), this toroidal structure would correspond to 
shocked gas in the equatorial region expanding perpendicularly to the jet \cite[see e.g.][]{Soker2000}. It is expected that 
the velocity of the shock 
would decrease as it interacts with the ambient gas of the CSE. Therefore, the value derived assuming a constant 
expansion velocity would be just an upper limit of the lifetime of the torus. As mentioned earlier, a complete 
understanding of the acceleration history of the gas is necessary to calculate the exact lifetime of the torus. In the 
case of a jet-driven bow shock the lifetimes of the toroidal structure and the jet should be the same.

It is worth to note that the location of the H$_{2}$O masers observed by \cite{Claussen2009} in the P-V diagram  
corresponds to emission from the LVC of the molecular gas. Therefore, it is unlikely that they are tracing the bow shock 
at the tip of the jet but rather they seem to be associated to the bow shock expanding laterally. This could explain the 
observed cease of expansion of the H$_{2}$O masers in this source \citep{Rogers2012}, although this could also be 
due  to the masers having reached the edge of the dense region of the CSE, as it is suggested by Fig.~\ref{Fig7}. Given this, 
we emphasise that the lifetime of the jet, hence the lifetime of the whole molecular outflow, cannot be obtained directly 
from the kinematical timescale of the H$_{2}$O masers. It is reasonable to expect that this is also true for other wf-nebulae, 
as discussed by \cite{Yung2017}. 

\subsection{The nature of the EHVO}\label{EHVO_section}

The molecular outflow considered in this work, which in the nomenclature of \cite{Sahai2017} is referred to as HVO, and the EHVO 
exhibit significant differences in their intrinsic properties. Firstly, as pointed out by \cite{Sahai2017}, there is an angle of $\sim$12$^{\circ}$ 
between the directions of the main axes of the EHVO and the HVO [see Fig.~\ref{Fig7} and Fig.~1 of \cite{Sahai2017}]. Secondly, 
considering the apparent extension of the EHVO on the plane of the sky ($\sim$$14^{\prime\prime}$); the projected radial velocity 
\citep[$\sim$310~km~s$^{-1}$;][]{Sahai2017}, and assuming constant velocity, the estimated kinematical timescale is 
$\tau_{\rm EHVO}$$\sim$215$\times$${\rm tan}\,\theta_{EHVO}$~years, 
where $\theta_{EHVO}$ is the inclination of the EHVO with respect to the plane of the sky. This timescale is larger than the one derived 
in the previous subsection for the HVO, although the actual value depends on the inclination, as well as on the initial velocity 
and the deceleration of the EHVO. Finally, the EHVO exhibits an expansion velocity that does not correspond to the velocity 
expected from the kinematical model presented in this work. We thus conclude that the clumps of the EHVO seen in Fig.~\ref{Fig7}, 
enclosed by an orange dashed line, correspond to material that was swept up by a different outflow, which is faster and likely older 
than the HVO. Additionally, there seems to be traces of entrained material of the EHVO outflow that appear as ``horns'' of the HVO 
and confined within the region traced by the CO($J$=1$\rightarrow$0) emission. Therefore, it is likely that the extension of the region traced by the 
CO($J$=1$\rightarrow$0) emission determines the radius of CO photodissociation by interstellar UV radiation.

\subsection{Kinetic energy and scalar momentum of the molecular outflow}

Since our model is fundamentally different from the one presented by \cite{Sahai2017}, here we recalculate the scalar 
momentum and kinematic energy of the molecular outflow. To obtain the kinetic energy and scalar momentum, in principle, 
it is possible to use the mass spectrum shown in Fig.~\ref{Fig6}. However, the velocity offset in this plot is just the 
component of the expansion velocity on the line-of-sight. The 3D expansion velocity is necessary for the calculations. 
To circumvent this problem, one can take advantage of the morphology adopted for the emitting region, which has an 
elongated narrow shape. In such a case, it is possible to de-project the velocity vectors by multiplying the velocity offsets 
by a factor 1/sin$\,\theta_{\rm inc}$, with $\theta_{\rm inc}$=45$^{\circ}$. This is not strictly true for all the velocity vectors, 
since the gas is not really moving along 
one single line. Particularly, for the gas located close to the central source, which might be expanding along the line-of-sight, 
the expansion velocity will be overestimated by a factor of 1/sin$\,\theta_{\rm inc}$. For such gas, this leads to an overestimation 
of the scalar momentum by a factor of $\sqrt{2}$ and for the kinetic energy by a factor of 2. In addition, there is an uncertainty 
due to the optical depth of the CO($J$=3$\rightarrow$2) emission since  it was assumed to be optically thin to obtain the mass 
spectrum of Fig.~\ref{Fig6}. If the emission is optically thick the estimated value of the mass represents just a lower limit. 
Having this in mind, we proceeded to calculate the kinetic energy and scalar momentum in the way introduced by 
\cite{Bujarrabal2001} and described in the following. Firstly we multiplied the mass in each channel by the square of the 
corresponding velocity to obtain the kinetic energy per channel. By adding the kinetic energy of all the channels we obtained 
$K_{\rm LVC}$=8$\times$10$^{44}$~erg and $K_{\rm HVC}$=1$\times$10$^{45}$~erg for 
the LVC and HVC, respectively. Similarly, the derived scalar momentum is 1.6$\times$10$^{38}$~gm~cm~s$^{-1}$ and 
6$\times$10$^{37}$~gm~cm~s$^{-1}$ for the LVC and HVC, respectively. The total kinetic energy and scalar momentum in 
the outflow are of the same order of magnitude as the values calculated by \cite{Sahai2017}. Subsequently, the scalar 
momentum in the molecular outflow transferred by the jet, $\Delta P$, is obtained by subtracting the scalar momentum due to the 
AGB wind to the total scalar momentum: $\Delta P$=$(P_{\rm LVC}+P_{\rm HVC})-(M_{\rm outflow}$$\times$${\rm v}_{\rm AGB})$, 
where $P_{\rm LVC}$ and $P_{\rm LVC}$ are the scalar momentum of the LVC and HVC, respectively; v$_{\rm AGB}$ is the 
expansion velocity of the AGB wind and $M_{\rm outflow}$ is the total mass of the outflow. From the mass spectrum in 
Fig.~\ref{Fig6} we estimate a total molecular mass in the outflow $M_{\rm outflow}$=1.6$\times$10$^{-2}$~$M_{\odot}$, thus 
$\Delta P$=1.6$\times$10$^{38}$~gm~cm~s$^{-1}$, where we have assumed an expansion velocity of the AGB wind 
v$_{\rm AGB}$$\sim$20~km~s$^{-1}$. As pointed out by \cite{Sahai2017}, given a star luminosity of 
$L_{\star}$=6000~$L_{\odot}$ for IRAS~16342$-$3814, the radiation alone cannot account for the scalar momentum of the 
molecular outflow. The mass-loss rate of the driving jet can be estimated by considering the extreme case in which the kinetic 
energy of the jet is purely transformed into kinetic energy of the molecular outflow: 
$\dot{M}_{\rm jet}$=[$K_{\rm LVC}$+$K_{\rm HVC}-0.5(M_{\rm outflow}$$\times$${\rm v}_{\rm AGB}^{2}$)]/(v$_{\rm jet}^{2} \tau_{\rm jet}$), 
where $K_{\rm LVC}$ and $K_{\rm LVC}$ are the kinetic energies of the LVC and HVC, respectively, and the kinetic 
energy due to the AGB wind has been subtracted. Using v$_{\rm jet}$=380~km~s$^{-1}$ and 
$\tau_{\rm jet}$=100~yr, the resulting mass-loss rate of the primary jet, i.e. the one driving the molecular outflow, is 
$\dot{M}_{\rm jet}$=5$\times$10$^{-6}$~$M_{\odot}$~yr$^{-1}$. As a matter of fact, it is known that part of the energy transferred 
by the jet is radiated away, e.g. H$_{2}$ emission \citep{Gledhill2012}, thus this value represents just a lower limit for the mass-loss 
rate of the jet.

\subsection{The precessing jet in IRAS~16342$-$3814}

The distribution of the emission peak positions of the high-velocity emission in Fig.~\ref{Fig1} a hints an oscillating pattern 
with respect to the direction of the major axis of the structure. To further investigate the distribution of the 
peak positions of the high-velocity emission we rotated the data points by an angle $\theta_{\rm rot}$$\approx$23$^{\circ}$ 
and plotted them as it is shown in Fig.~\ref{Fig8}. From this plot it is clear that the pattern of the peak positions exhibits 
an oscillating point-symmetric pattern.

This pattern can be explained by the presence of a collimated outflow that precesses around a fixed axis. From 
Fig.~\ref{Fig8} it can be seen that the amplitude of the oscillations is much smaller than extent of the pattern, 
thus implying a very small precessing angle. If we assume that every particles of the jet has 
decelerated at a constant rate, $a_{n}$=$[{\rm v}(r_{\rm n})^{2}-{\rm v}_{\rm i}^{2}]/(2 \, r_{\rm n}$), during its journey from 
the base of the outflow to the radius $r$=$r_{\rm n}$, it is possible to express the oscillating pattern of the emission peak 
positions with a sinusoidal function expressed as: 
\begin{equation}\label{Eq:2}
y(r)=A\times r \times {\rm sin}\left[\frac{2 \, \pi \, \tau(r)}{T}-\phi \right], 
\end{equation}
where $\tau(r)$ is the travelling-time of the particle from the base of the outflow to the radius $r$ and it is given by 
\begin{equation}\label{Eq:3}
\tau(r)=-{\rm v}_{\rm i}+\frac{\left[{\rm v}_{\rm i}^{2}+2 \, r \, a(r)\right]^{1/2}}{a(r)}, 
\end{equation}
where v$_{\rm i}$ is the initial velocity of the jet and $a(r)$ is the acceleration of the particle of the jet located at a distance 
$r$. The factor $A$$\times$$r$ in Equation \ref{Eq:2} expresses the amplitude of the sinusoidal pattern as a function of $r$. 
Thus, the parameter $A$ is the tangent of the semi-opening angle of the pattern. $T$ is the precession period and 
$\phi$ is the phase of the sinusoidal pattern. 

We fitted the precession model given by Equation \ref{Eq:2} to the data points of the pattern seen in Fig.~\ref{Fig8}. 
However, the shape of the pattern depends on the angle, $\theta_{\rm rot}$, used for rotating the positions 
of the data points. Therefore, $\theta_{\rm rot}$ is also a free parameter. Considering the P.A. of the line connecting the 
H$_{2}$O masers detected by \citep{Claussen2009}, the rotation angle is $\theta_{\rm rot}$=23$\rlap{.}^{\circ}$9 and the 
best fit parameters are $A$=0.018$\pm$0.001, $T$=86$\pm$4~yrs, $\phi$=57$\pm$10$^{\circ}$. If we adopt the P.A. 
of the nebula's long axis determined by \citep{Sahai2017}, the rotation angle is $\theta_{\rm rot}$=23$^{\circ}$, which 
results in the following best fit parameters: $A$=0.010$\pm$0.001, 
$T$=58$\pm$5~yrs, $\phi$=29$\pm$30$^{\circ}$. In Fig.~\ref{Fig8} we plot the best fit function for an intermediate 
value of the rotation angle angle $\theta_{\rm rot}$=23$\rlap{.}^{\circ}$5. The blue line in the plot shows the fit to the 
data with parameters, $A$=0.014$\pm$0.001, $T$=72$\pm$4~yrs, $\phi$=$-$17$\pm$16$^{\circ}$. The corresponding 
opening angle of the pattern can be obtained as 2$\times$atan($A$)=$\theta_{\rm op}$$\approx$2$^{\circ}$. Given the 
limited angular and spectral resolution of these observations, it is difficult to discern which rotation angle gives the best 
fit. Higher angular and spectral resolution observations are necessary to unambiguously identify the symmetry axis of 
the jet and better constrain its parameters.  

The precession period derived for the sinusoidal pattern of the CO emission peak positions is slightly longer than the 
one obtained for the corkscrew pattern found by \cite{Sahai2005} from IR observations, $T$$\lesssim$50~years, 
even considering the uncertainty in the rotation angle. The period of the corkscrew pattern was 
obtained assuming that the velocity of the dust is similar to the velocity of the OH masers, 
v$_{\rm OH}$$\sim$90~km~s$^{-1}$. However, under the action of an external force, such as the one exerted by the 
ram pressure of the driving jet, the velocity of the dust could be lower that the velocity of the gas, since the inertial mass 
of the dust grains is larger than the one of the gas particles. Indeed, \cite{Yoshida2011} found that the dust in the superwind 
of the starburst galaxy M82 is kinematically decoupled from the gas, with the former moving substantially slower than the latter. 
In such a case, the value obtained by \cite{Sahai2005} would represent a lower limit for the period. 
For example, if one takes an expansion velocity of v$_{\rm exp}$=50~km~s$^{-1}$ for the dust, the period of the corkscrew 
pattern is $\sim$70~years, which is within the range of the values obtained for the CO pattern. We propose that the 
sinusoidal pattern seen in Fig.~\ref{Fig8} traces the underlying outflow that drives the dust outwards and forms the 
corkscrew pattern seen in the IR. 

\section{Conclusions}

In this work we have analysed the CO($J$=3$\rightarrow$2) emission of the water fountain nebula IRAS~16342$-$3814 to 
further study the spatio-kinematical characteristics of the molecular gas. We identified two components with different 
kinematical characteristics; one that exhibits deceleration, and the other with the velocity increasing as a function of 
distance. We proposed a spatio-kinematical model that includes a decelerating collimated component surrounded by 
more extended component with a positive velocity gradient. The elongated narrow morphology of the model is in 
good agreement with results from previous IR and H$_{2}$O maser observations. We created synthetic images and 
P-V diagrams using the morphology and velocity fields of our model. The resulting synthetic images and P-V diagrams 
reproduce very well the observations. The morphology and velocity field, together with the mass spectrum derived 
from the emission, of the gas are consistent with a jet-driven molecular outflow. The decelerating component 
is associated to material entrained by the underlying jet near its axis or it could be the molecular component of the jet itself. 
This feature is not seen in other more evolved objects that exhibit more developed bipolar morphologies. We conclude 
that it is likely that the jet in those objects has already disappeared as it is expected to last only a couple of hundred years. 
This strengthens the idea that water fountain nebula are undergoing a very short transition where they develop the 
collimated outflows that shape the CSE. The timescale for the molecular outflow is $\tau_{\rm jet}$$\sim$70-100~years. 
The scalar momentum carried by the molecular outflow is much larger than it can be provided by the radiation of the 
central star during the short lifetime of the outflow. We identified an oscillating point-symmetric pattern for the 
component. We fitted the data points using a sinusoidal function and derived a period of $T$$\sim$60-90~years and an 
opening angle of $\theta_{\rm op}$$\approx$2$^{\circ}$. We propose that this pattern traces the jet that drives the 
material that forms the corkscrew pattern seen in the IR.

 ALMA is allowing for the first time to study in full detail the morphology of collimated molecular outflows underlying the 
 high-velocity structures traced by H$_{2}$O maser in wf-nebulae. A follow-up study of these objects with ALMA 
 would be of great importance to understand the launching, collimation and interaction with the circumstellar material 
 of the collimated outflows of wf-nebuale.

\begin{acknowledgements}
This paper makes use of the following ALMA data: ADS/JAO.ALMA\#2012.1.00678.S. ALMA is a partnership of 
ESO (representing its member states), NSF (USA) and NINS (Japan), together with NRC (Canada), MOST and ASIAA 
(Taiwan), and KASI (Republic of Korea), in cooperation with the Republic of Chile. The Joint ALMA Observatory is 
operated by ESO, AUI/NRAO and NAOJ. DT was supported by the ERC consolidator grant 614264. GO would like to 
acknowledge financial support from the NSFC (Grant No. 11503072) and the National Key R\&D Program of China 
(Grant No. 2018YFA0404602), and the Youth Innovation Promotion Association of the CAS. The authors thank Susanne 
Aalto for helpful discussions. The authors also thank the anonymous referee for constructive comments and suggestions 
that helped to improve the manuscript.
\end{acknowledgements}

%
%

%
%

\bibliography{bibliography_tafoya} 

\end{document}